# Analysis and Study of Incremental DBSCAN Clustering Algorithm

| | |
|---|---|
| SANJAY CHAKRABORTY | Prof. N.K.NAGWANI |
| National Institute of Technology | National Institute of Technology |
| (NIT) Raipur, CG, India. | (NIT) Raipur, CG, India. |

**ABSTRACT**

This paper describes the incremental behaviours of Density based clustering. It specially focuses on the Density Based Spatial Clustering of Applications with Noise (DBSCAN) algorithm and its incremental approach.DBSCAN relies on a density based notion of clusters.It discovers clusters of arbitrary shapes in spatial databases with noise.In incremental approach, the DBSCAN algorithm is applied to a dynamic database where the data may be frequently updated. After insertions or deletions to the dynamic database, the clustering discovered by DBSCAN has to be updated. And we measure the new cluster by directly compute the new data entering into the existing clusters instead of rerunning the algorithm.It finally discovers new updated clusters and outliers as well.Thus it describes at what percent of delta change in the original database the actual and incremental DBSCAN algorithms behave like same.DBSCAN is widely used in those situations where large multidimensional databases are maintained such as Data Warehouse.

**Keywords**
Clustering, Data mining, DBSCAN, Density, Distance-connectivity, Distance-reachability, eps, Incremental, Minpts, Outlier.

## 1. INTRODUCTION

### 1.1 Data Mining

Data mining, as one of the promising technologies since 1990s, is some to extent a non traditional data driven method to discover novel, useful, hidden knowledge from massive data sets. Data clustering is the most famous and necessary concepts in data mining.

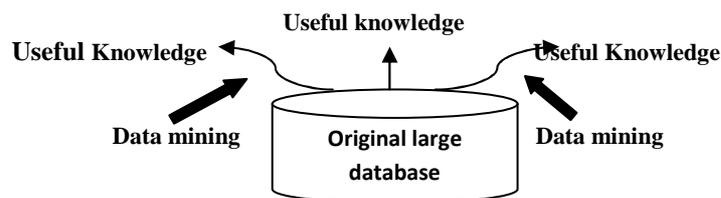

**Figure 1: Extraction of knowledge from a large database**

### 1.2 Incremental Data Mining

Incremental data mining means applying Data mining Algorithms on incremental Database. The objective of incremental data mining algorithms is to minimize the scanning and calculation effort for newly added records. Here we improve the efficiency of newly added record updating problem. These are the few prime factors that cause to apply the incremental Mining.

 i. Database always gets modified.
ii. During each modification (insertion and deletion) the database requires scanning again. Thus it requires lot of time for rescanning.
iii. Hence, the Incremental version of clustering algorithm includes the logic for Insertion and Deletion as separate Dynamic operation.
iv. Whenever, the databases may have frequent updates and thus it may be dynamic.
v. After insertions and deletions to the database, the existing clustering algorithm has to be updated.
vi. After that we expect algorithm of incremental approach performs efficiently compare to actual one.

### 1.3 Clustering

Data clustering is a popular unsupervised clustering approach for automatically finding classes, concepts, or group of patterns. Clustering involves dividing a set of objects into a specified number of clusters. The process of grouping a set of physical or abstract objects into classes of similar objects is called clustering [9]. There are many types of data that often occur in cluster analysis such as interval-scaled variables, binary variables, nominal, ordinal, mixed and ratio variables. Clustering is a challenging field of research in which its potential applications pose their own special requirements.In this paper,the incremental clustering algorithm is represented. This algorithm is based on DBSCAN which is an efficient clustering algorithm for metric databases (that is, databases with a distance function for pairs of objects) for mining a large database.

DBSCAN algorithm is particularly suited to deal with large datasets, with noise, and is able to identify clusters with different sizes and shapes. Density based clustering has proven to be very effective for analyzing large amounts of heterogeneous, complex data for example clustering of complex objects [1]. This paper discuss about actual DBSCAN and Incremental DBSCAN clustering from data mining perspective. The main inherent idea of this paper is to compare the percent of time needed to run the whole DBSCAN algorithm with respect to insertion of new coming data objects and determines which approach takes fewer amounts of time and effort. Due to the density-based nature of DBSCAN, the insertion or deletion of an object affects the current clustering only in the neighborhood of this object.

### 1.4 DBSCAN clustering

One of the most common clustering algorithms and also most cited in scientific literature is Density Based Spatial Clustering of Applications with Noise (DBSCAN) which has the ability to produce arbitrary shape of clusters.Clusters are identified by looking at the density of points. Regions with a high density of points depict the existence of clusters whereas regions with a low density of points indicate clusters of noise or clusters of outliers. DBSCAN grows clusters according to a density based connectivity analysis. It defines a cluster as a maximal set of density-connected points. The key idea of density-based clustering is that for each object of a cluster the neighborhood of a given radius (eps) has to contain at least a minimum number of objects (MinPts), i.e. the cardinality of the neighborhood has to exceed some threshold [2].

The algorithm DBSCAN was designed to efficiently discover the clusters and the noise in a database according to the above definitions. DBSCAN requires two parameters: $\varepsilon$ (eps) and the minimum number of points required to form a cluster (Minpts). It starts with an arbitrary starting point that has not been visited. This point's $\varepsilon$-neighborhood is retrieved, and if it contains sufficiently many points, a cluster is started. Otherwise, the point is labeled as noise.

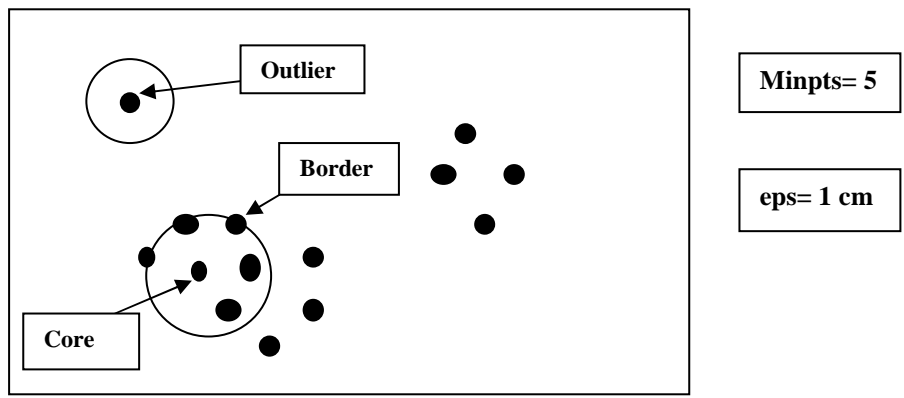

**Figure 2: Clustering of a set of objects based on the DBSCAN**

**1.5 Paper Organization**

The rest of this paper is organized as follows. Section 2 discusses Literature Survey on clustering techniques. The formulas and proposed algorithms for incremental DBSCAN clustering are presented in Section 3 and proposed model and Illustrative examples are reported in section 3.2 & 3.3 respectively. Section 3.4 describes the benefits of DBSCAN

clustering.Section 4 concludes with a summary of those clustering techniques and Future scope of this work. Section 5 describes the references.

## 2. LITERATURE SURVEY

There are number of clustering techniques but this paper mainly focuses on the widely used DBSCAN clustering technique. Significant work is done in the field of Density based clustering. This section represents a brief overview of some work done in Density based clustering including email classification,spam detection etc.

Number of application areas and techniques are highlighted in Density-based clustering.One approach developed the incremental clustering for mining large database environment.This approach present the first incremental clustering algorithm based on DBSCAN clustering which is applicable to any database containing data from a metric space. Due to the density-based nature of DBSCAN, the insertion or deletion of an object affects the current clustering only in the neighbourhood of this object. Thus, efficient algorithm scan be given for incremental insertions and deletions to an existing clustering.Incremental DBSCAN yields significant speed-up factors over DBSCAN even for large numbers of daily updates in a data warehouse. In this paper, sets of updates are processed one at a time without considering the relationships between the single updates [1].

A innovative technique which is used to compare in between two different clustering algorithms (DBSCAN and SNN) described several implementations of the DBSCAN and SNN algorithms, two density-based clustering algorithms. These implementations can be used to cluster sets of points based on their spatial density. The results obtained through the use of these algorithms show that SNN performs better than DBSCAN since it can detect clusters with different densities while DBSCAN cannot [3].

Many clustering algorithms have been proposed so far, seldom was focused on high-dimensional and incremental databases. An incremental approach on Grid Density-Based Clustering Algorithm (GDCA) discovers clusters with arbitrary shape in spatial databases. It first partitions the data space into a number of units, and then deals with units instead of points. Only those units with the density no less than a given minimum density threshold are useful in extending clusters. An incremental clustering algorithm--IGDCA is also presented in this paper, applicable in periodically incremental environment [10].

An innovative approach presents a new density-based clustering algorithm, ST-DBSCAN, which is based on DBSCAN. It proposes three marginal extensions to DBSCAN related with the identification of (i) core objects, (ii) noise objects, and (iii) adjacent clusters. In contrast to the existing density-based clustering algorithms, this algorithm has the ability of discovering clusters according to non-spatial, spatial and temporal values of the objects. In this paper, it also presents a spatial–temporal data warehouse system designed for storing and clustering a wide range of spatial–temporal data [9].

One of the new concepts is presented on clustering technique which provides an effective method for Clustering Incremental Gene Expression data. It is designed based on density based approach where the efficiency of GenClus in detecting quality clusters over gene expression data. This work presents a density based clustering approach which finds useful subgroups of highly coherent genes within a cluster and obtains a hierarchical structure of the dataset where the sub-clusters give the finer clustering of the dataset[2].

In some cases, a fast incremental clustering algorithm is used to handle dynamic databases. It has the ability to changing the radius threshold value dynamically. The algorithm restricts the number of the final clusters and reads the original dataset only once. At the same time the frequency information of the attribute values is introduced by this algorithm. It can be used for the categorical data. The algorithm can not only overcome the impact of the inadequate of the memory when clustering the large scale data set, but also accurately reflect the characteristics of the data set [14].

A famous concept which discovers distributed clustering environment using DBSCAN algorithm.This approach ultimately forms a distributed clustering algorithm.This approach helps to cluster the data locally and independently from each other and transmitted only aggregated information about the local data to a central server [15].

Therefore incremental DBSCAN clustering algorithm requires less space and less time than the actual one.

## 3. INCREMENTAL DBSCAN CLUSTERING

The term incremental means "% of $\delta$ change in the original database" i.e. insertion of some new data items into the already existing clusters. Such as,

$$\% \delta \text{ change in DB} = \frac{(NEW\ DATA - OLD\ DATA)}{OLD\ DATA} \times 100 \qquad [1]$$

Sometimes DBSCAN may be applied on dynamic database which is frequently updated by insertion or deletion of data. After insertions and deletions to the database, the clustering discovered by DBSCAN has to be updated. Incremental clustering could improve the chances of finding the global optimum. In this approach, first we form clusters based on the initial objects and a given radius(eps) and minimum number of points(Minpts).Thus we finally get some clusters fulfilling the conditions and some outliers.Now, when new data are inserting into the existing database, we have to update our clusters using DBSCAN.At first we compute the means between every core object of clusters and the new coming data and insert the new data into a particular cluster based on the minimum mean distance.The new data which are not inserted into any clusters,they are treated as noise or outliers.Sometimes outliers which fulfil the Minpts & eps criteria,combinly can form clusters using DBSCAN.

### 3.1 Proposed Algorithm

The following steps describe the proposed algorithm which points out the comparison of actual and incremental K-means clustering.

---
**Input:**
D: A dataset containing n objects.{$X_1, X_2, X_3 ..., X_n$}
n: number of data items.
Minpts: Minimum number of data objects.

---

eps: radius of cluster.
**Output:**
$K_1$: A Set of clusters.
**Algorithm:**
Let, $C_i$ (where i=1, 2, 3 …) is the new data item.
**1.** Run the actual DBSCAN algorithm and clustered the new data item Ci properly based on the radius(eps) and the Minpts criteria. Repeat till all data items are clustered. Actual DBSCAN ⟶ $T_1$ (Processing Time).

**Incremental DBSCAN Pseudo-code:**
**Start:**
**2.** a> Let, K represents the already existing clusters.
  b>When new data is coming into the database, the new data will be directly clustered by calculating the minimum mean(M) between that data and the core objects of existing clusters.
for i = 1 to n do
find some mean M in some cluster Kp in
K such that dis ( $C_i$, M ) is the smallest;
If (dis ( $C_i$, M ) is minimum) && ($C_i$<=eps) && (size(Kp)>=Minpts) then
Kp = Kp U $C_i$ ;
Else
If dis ($C_i$ != min) || ($C_i$>eps) ||(size(Kp)<Minpts) then $C_i$ ⟶ Outlier($O_i$)        .
Else
If Count($O_i$)>= Minpts then $O_i$ ⟶ Form new cluster($M_i$).
**C >** Repeat step b till all the data samples are clustered.
 Incremental DBSCAN ⟶ $T_2$. (Processing Time).
**End ;**
 **3.** Compare ($T_1$, $T_2$)
   Result: ($T_2$< $T_1$) ⟵ Upto some certain Threshold value in the database.

> 4. Compute the actual threshold value.
>    Finds the incremental clustering as a better time efficient approach.

### 3.2 Proposed Model

Problem: In dynamic databases, if new transactions (records/rows) are appended as time advances. It is an incremental algorithm, used to deal with this problem. The proposed algorithm identifies the value of percentage of size of original database x, which can be added to original database. Now there might be two following cases:

1. Up to x% change in the original database, better to use previous result.
2. For more than x %, rerun the algorithm again.

Solution: As a result, Incremented DBSCAN will provide faster execution than the algorithms used previously DBSCAN because the number of scans for the database will be decreased.

The proposed model can be explained using figure-3 which defines the actual DBSCAN algorithm is applied on the original database and store the result into another database using mysql or others. Now the incremental DBSCAN algorithm is applied on the incremental dataset where the results of these two are compared and also evaluate the performance.

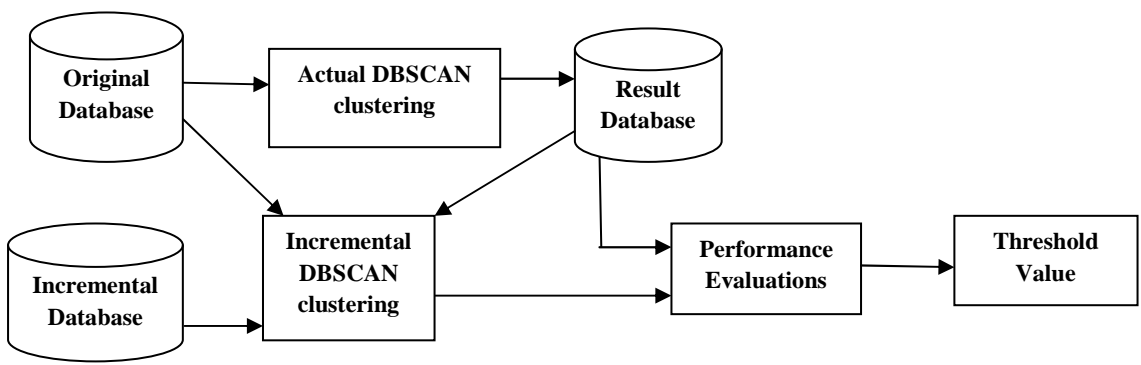

Figure 3: Performance evaluations of incremental DBSCAN clustering

### 3.3 Illustrative Examples of Proposed Model
The proposed model can be explained with the help of following two examples:

**Example.1**. Suppose there is a set of data objects, such as12,15,22,85,82,73,8,10,17,48,96,152.Initially assume there are two clusters.Also assume that the points 15 and 82 are core objects of the two clusters. Form clusters properly using DBSCAN algorithm.Suppose two new data 56 and 125 are inserted into the existing database.Then show how this algorithm will behave? (eps=1cm and Minpts=5)

`Ans:` Assume that there are initially two clusters, where 15=core object of the first cluster and 82= core object of the second cluster. Now based on the radius (eps) and minimum number of objects(Minpts),we can get

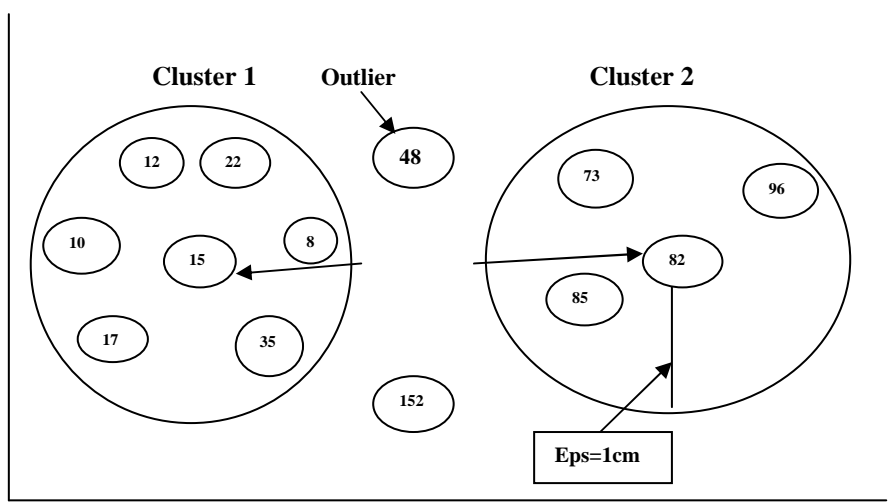

Now,when a new data is coming such as data=56, the new data will be directly clustered by calculating the minimum mean(M) between that data and the core objects of existing clusters.There is no need to rerun the whole DBSCAN algorithm again.So here after using Manhattan distance metric, Cluster1=|56-15|=41

Cluster2=|82-56|=26(minimum), then 56 is clustered into the Cluster2 as it satisfy that dis(56,82)<=eps. So the new data item should be entered into the 'Cluster 2' directly without rerunning the whole algorithm. Thus it saves our time and effort both.

Now another data 125 is inserted then Cluster1=|125-15|=110

Cluster2=|125-86|=43(minimum), then 56 is clustered into the Cluster2 but dis(125,82)>eps, so the data 125 is now treated as outlier or noisy data.

**Example2:** Suppose there is a set of two dimensional data objects, such as (5,2), (12,3),(22,82),(125,110),(32,42),(12,28),(56,48) and (68,72).Initially assume there are two clusters.Also assume that the points (12,3) and (32,42) are core objects of the two clusters. Form clusters properly using DBSCAN algorithm.Suppose four new data (132,122),(38,58),(162,135) and (118,126) are inserted into the existing database.Then show how this algorithm will behave? (eps=3cm and Minpts=4)

**Ans:** Assume that there are initially two clusters, where (12,3) =core object of the first cluster and (32,42) = core object of the second cluster. Now based on the radius (eps) and minimum number of objects(Minpts),we can get

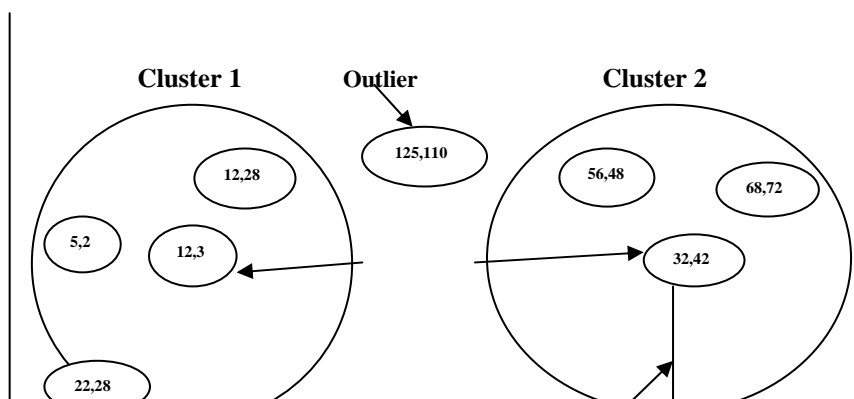

**Core data**

New data= (132,122),(38,58),(162,135) and (118,126)

So, 1. (132,122)>eps of any cluster, then it will treated as outlier.

2. (38,58) => Cluster1=|(38-12)+(58-3)| =81

Cluster2=|(38-32)+(58-42)|=22(minimum), then (38,58) is clustered into the Cluster2 as it satisfy that dis[(38,58) ,(32,42)]<=eps. So the new data item should be entered into the 'Cluster 2' directly without rerunning the whole algorithm. Thus it saves our time and effort both.

3. (162,135) and (118,126) both of them are >=eps, so both of them are treated as outliers.

Now a new concept is aroused that these three outliers form a new cluster with the old outlier (125,110). Thus, the new cluster is Cluser3={(162,135), (118,126), (125,110), (132,122)}.

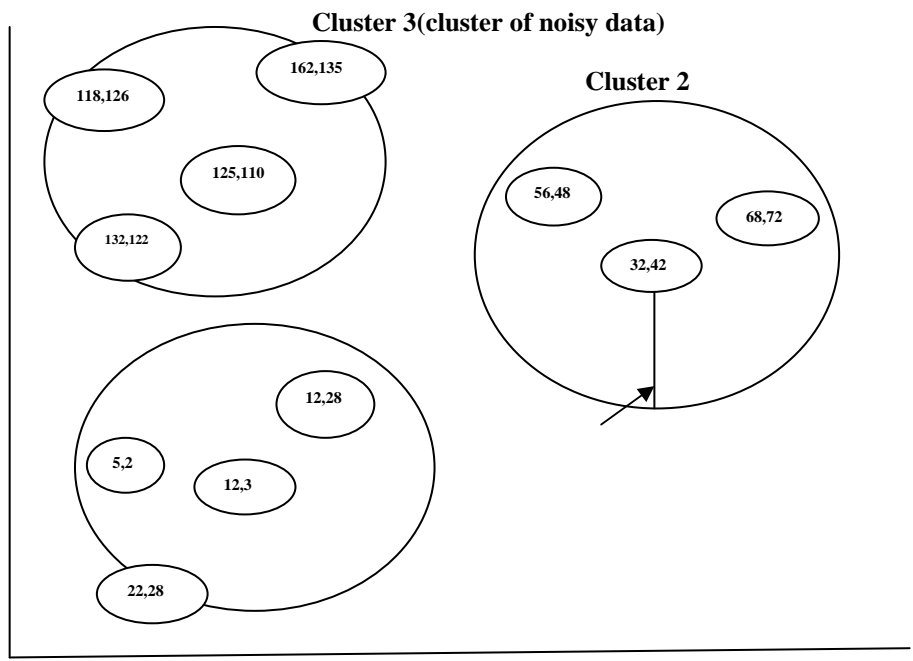

Eps=3cm

Cluster 1

### 3.4 Benefits of Incremental DBSCAN Algorithm
The actual DBSCAN approach is not suitable for a large multidimensional database which is frequently updated. In that case, the incremental clustering approach is much better. The system with incremental concept saves lot of time and effort efficiently whereas the existing system has already suffering with some drawbacks and these problems are mainly faced in dynamic large databases by the existing system. When some records are added to existing data, then it deals with this problem of scanning the whole database again. The time complexity is very high due to rescanning the whole database. It requires more effort as well. The Existing system is not efficient with respect to time and effort. That's why new system with incremental clustering approach is more suitable to use in a large multidimensional dynamic database.

### 4. CONCLUSION AND FUTURE SCOPE
This paper propose an improve DBSCAN clustering approach which provides better and fastest result compare to the existing DBSCAN clustering algorithm up to some certain point of change in the original database. In this paper the efficiency of the incremental DBSCAN clustering algorithm is estimated. The future scope of the work could be analyzing the other popular clustering techniques in incremental fashion.